\documentclass[aps,prd,reprint,groupedaddress,amssymb,amsmath]{revtex4-1}
\usepackage{hyperref}
\usepackage{bm}
\usepackage{makeidx}
\usepackage{amsmath}
\usepackage{graphicx}
\usepackage{dcolumn}
\usepackage{bm}
\usepackage{color}
\usepackage{amsfonts}
\usepackage{amssymb}

\begin{document}

\title{Magnon excitations and quantum critical behavior of the ferromagnet U$_4$Ru$_7$Ge$_6$}

\author{M. P. Nascimento$^{1}$}
\author{ M. A. Continentino$^{1}$}
\author{A. L\'opez$^{2}$}
\author{Ana de Leo$^{2}$}
\author{D. C. Freitas$^{3}$}
\author{J. Larrea J.$^{4}$}
\author{Carsten Enderlein$^{5}$}
\author{J. F. Oliveira$^{1}$}
\author{E. Baggio-Saitovitch$^{1}$}
\author{Jir\'i Posp\'isil$^{6}$}
\author{ M. B. Fontes$^{1}$}

\affiliation{$^{1}$Centro Brasileiro de Pesquisas F\'{\i}sicas, 22290-180, Rio de Janeiro , Brazil}
\affiliation{$^{2}$Departamento de F\'{\i}sica Te\'orica, Universidade do Estado do Rio de Janeiro,  20550-013,
Rio de Janeiro, RJ, Brazil.}
\affiliation{$^{3}$Instituto de F\'{\i}sica, Universidade Federal Fluminense, 24210-346, Niter\'oi, RJ, Brazil}
\affiliation{$^{4}$Instituto de F\'{\i}sica, Universidade de S\~ao Paulo, 05508-090, S\~ao Paulo, SP, Brazil.}
\affiliation{$^{5}$Universidade Federal do Rio de Janeiro, Rio de Janeiro-RJ, Brazil}
\affiliation{$^{6}$Faculty of Mathematics and Physics, Charles University, DCMP, Ke Karlovu 5, CZ-12116 Praha 2, Czech Republic}

\date{\today}

\begin{abstract}
We present an extensive study of the ferromagnetic heavy fermion compound U$_4$Ru$_7$Ge$_6$. Measurements of electrical resistivity, specific heat and magnetic properties show that U$_4$Ru$_7$Ge$_6$ orders ferromagnetically at ambient pressure with a Curie temperature $T_{C} = 6.8 \pm 0.3$ K. The low temperature magnetic behavior of this soft ferromagnet is dominated by the excitation of gapless spin-wave modes. Our results on the transport  properties of U$_4$Ru$_7$Ge$_6$ under pressures up to $2.49$ GPa suggest that U$_4$Ru$_7$Ge$_6$ has a putative ferromagnetic quantum critical point (QCP) at $P_c \approx 1.7 \pm 0.02$ GPa.
In the ordered phase, ferromagnetic magnons scatter the conduction electrons  and give rise to a well defined power law temperature dependence in the resistivity. The coefficient of this term is related to the spin-wave stiffness  and measurements of the very low temperature  resistivity allow to accompany the behavior of this quantity as the the ferromagnetic QCP is approached. We find that the spin-wave stiffness decreases with increasing pressure implying that the transition to the non-magnetic Fermi liquid state is driven by the softening of  the magnons. The observed quantum critical behavior of the magnetic stiffness  is consistent with the influence of disorder in our system. At quantum criticality ($P = P_c \approx 1.7 \pm 0.02$ GPa), the  resistivity shows the behavior expected for an itinerant metallic system   near a ferromagnetic  QCP.

\end{abstract}


\maketitle


\section{Introduction}

The problems related to strongly correlated electronic systems are of great current interest due to the novel states of matter that can arise in these systems~\cite{0,01}. These include their exotic magnetic properties, superconducting behavior and their phase diagrams, which  exhibit quantum critical points (QCPs)~\cite{02}. QCPs are experimentally explored by doping, applied pressure or magnetic field~\cite{0}. In the case of  actinide materials, the interesting properties arise from partially filled $f$-orbitals that strongly hybridize with the conduction electrons. This, together with the strong correlations among the $f$-states give rise to a variety of ground states.

The ternary compound U$_4$Ru$_7$Ge$_6$ is a system with interesting magnetic properties. It has   a centered body (bcc) crystalline structure of the type U$_4$Re$_7$Si$_6$ \cite{1}. The lattice parameter is $a = 8.287$ \AA,   and the interatomic space between the Uranium is $d_{U-U} = 5.864$ \AA \cite{2}, much larger than the Hill boundary for Uranium: $d_{U-U} = 3.4$ $-$ $3.6$ \AA, which sets conditions for a  magnetic ground state~\cite{jirka1}. The compound U$_4$Ru$_7$Ge$_6$ has the properties of a heavy fermion system, with a Kondo resistivity and a large linear term in the low temperature specific heat~\cite{2}. It orders ferromagnetically at low temperatures due to the small volume of its unit cell, which favors the RKKY interaction~\cite{2, 3, 4}. Its ferromagnetism is characterized as itinerant, although Mentink et {\it al}.~\cite{2}  propose localized ferromagnetism, contrary to other works in the literature~\cite{5, 6, 7}. Under applied pressure, transport measurements show no evidence of discontinuous behavior as the ferromagnetic phase is suppressed and a non-magnetic Fermi liquid state is attained~\cite{6}.

In this work, we present an extensive study of the magnetic, thermodynamic and transport properties of U$_4$Ru$_7$Ge$_6$~\cite{7} under applied magnetic field and of the latter under high applied pressures. We show that  this system below its  ambient pressure ferromagnetic transition at $T_{C} = 6.8 \pm 0.3$ K has its low temperature properties dominated by the presence of low energy spin-wave excitations. In systems with strong magneto-crystalline anisotropy as compounds containing $f$-states, these modes are generally quenched at low temperatures by the existence of a gap in the spin-wave spectrum due to this anisotropy. However, U$_4$Ru$_7$Ge$_6$ is a unique system among the actinide materials with a negligible magneto-crystalline anisotropy~\cite{7}.
This allows for the excitation of magnons at very low temperatures  and these modes end up having a prominent role on  its low temperature physical properties, as we show here. In particular magnons scatter the conduction electrons and have a definite importance in the electrical resistivity of ferromagnetic metals. As we apply pressure on U$_4$Ru$_7$Ge$_6$ and measure its resistivity, we have a rare opportunity to observe the evolution of the spin-wave stiffness of a ferromagnetic system as it approaches the QCP. Our transport measurements show a clear softening of the magnon modes as U$_4$Ru$_7$Ge$_6$ is driven to the putative ferromagnetic QCP (FQCP) with increasing pressure.  In the present study, we obtain the quantum critical behavior of the stiffness of these excitations.

\section{Experimental}

The sample was prepared by arc melting  of its  high purity metallic
constituents in the ratio U:Ru:Ge = 1:2:2, under argon atmosphere,
without further heat treatment, as to form the compound URu$_2$Ge$_2$.
However, the X-ray diffractogram at room temperature showed a
composition of U$_4$Ru$_7$Ge$_6$,  with additional spurious phases,
as discussed in detail below. The X-ray powder diffraction (XRD)  at
room temperature was performed using the 
Bruker AXS D8 Advance II diffractometer, with LynxEye detector, Cu source with 
K$_{\alpha }$ radiation.

The
diffraction pattern was collected in a Bragg–-Brentano configuration
covering the angular range of $10$ to $90$ degrees, each step
incremented by $0.02$ degree. The XRD data were refined using the
Rietveld method~\cite{8}, implemented in the program FullProf~
\cite{9}, available  at the {\it Institut Laue-Langevin} (ILL)
website~\cite{10}. The profile function used to adjust the shape of
the diffraction peaks was the pseudo-Voigt function.

Pressure-dependent resistivity measurements were carried out in a temperature range from 0.1 K  to 10 K in a non-commercial Adiabatic Demagnetization Refrigerator (ADR). 
A standard 2.5 GPa piston cylinder type of cell was used, with a mixture of fluorinert FC70-FC77 as pressure medium, pure lead as pressure sensor at low temperatures and manganin as a manometer for loading the cell at ambient temperature.
We further performed measurements of electrical resistivity under magnetic fields up to 9 T and respective magnetoresistance measurements in the temperature range from 1.8 K to 30 K in a commercial PPMS Dynacool from Quantum Design, at ambient pressure.

The specific heat measurements as a function of temperature were also performed in the PPMS DynaCool  under different magnetic field values ranging from   0 to 7 T, in the temperature interval from 2 K to 15 K.

The magnetic characterization involved the application of external
magnetic fields in  DC and AC modes. The DC magnetization
measurements were in  field cooling (FC)  mode in a field of 10 mT. For the AC
susceptibility measurement, the parameters used  were $H_{AC} = 1$ 
mT and $H_{DC} = 50$ mT at 3 kHz. Both type of measurements were
performed from 2 K to 300 K in PPMS DynaCool,  Quantum Design.

\section{Sample analysis}

The XRD showed that the sample produced has a main phase of cubic
crystalline structure of   U$_4$Ru$_7$Ge$_6$ and secondary
phases. After a detailed analysis of the diffraction pattern and
identification of all peaks of minor intensities, it was found that
the secondary phases could probably be Ru$_2$Ge$_3$ and $\gamma$-U.
The Rietveld refinement of the X-ray diffractogram was performed
using first the U$_4$Ru$_7$Ge$_6$ main compound phase. In the
sequence, the CIF data of the other phases were added in the base of
the program; after adjustment, the presence of Ru$_2$Ge$_3$ and
$\gamma$-U compounds was confirmed as secondary phases in the
sample. The CIF data were obtained from the Inorganic Crystal Structure Database (ICSD).

\begin{figure}[!h]
\begin{center}
\includegraphics[scale=0.4]{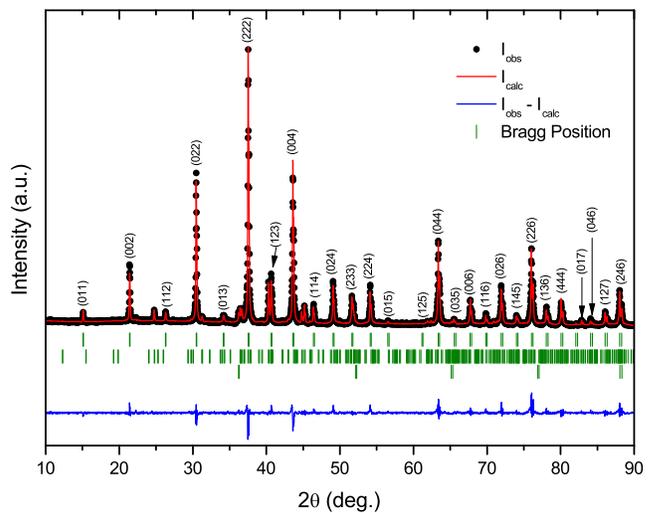}
\end{center}
\caption{(Color online) Result of the Rietveld refinement of the
X-ray diffraction pattern of the sample. The experimentally observed
data are the black circles and the calculated standard is the solid
red line. The allowed Bragg positions are represented in vertical
green bars, where each level corresponds to each of the phases
found: U$_4$Ru$_7$Ge$_6$ (78.06\%), Ru$_2$Ge$_3$ (20.78\%) and
$\gamma$-U (1.16\%). The difference pattern ($I_{obs}$ - $I_{calc}$)
is represented in solid blue line. The planes (hkl) shown are of the
major phase U$_4$Ru$_7$Ge$_6$. } \label{Fig1}
\end{figure}

Fig.~\ref{Fig1} shows the refinement of the XRD, where the black
circles are the experimentally observed data, while the solid red
line is the standard calculated by the refinement. The allowed Bragg
positions are represented in vertical green bars, where each level
corresponds to the peaks of the Bragg planes of each of the phases
found. The planes (hkl) shown in the figure correspond to the peaks
of the diffractogram of the predominant phase U$_4$Ru$_7$Ge$_6$,
which appears in the amount of 78.08\%. The parameters of the
crystalline lattice and the amounts of each of the phases found as
results of the refinement, as well as their quality R factors, are
described in Table 1. The crystallographic data-sheet ICSD 192067~\cite{5} was used for the refinement of the phase U$_4$Ru$_7$Ge$_6$.
In Table 1, it is seen that the lattice parameters of all phases are
in agreement with the literature~\cite{2, 3, 12, 13}. The
interatomic distance of the Uranium of the major phase
U$_4$Ru$_7$Ge$_6$ is d$_{U-U}$ = 5.866 \AA, also agreeing with the
literature~\cite{2, 3}.

While  Uranium in its allotropic
forms has a higher resistivity than copper \cite{16}, it has a weak
paramagnetic behavior, exhibiting paramagnetism almost independent
of temperature \cite{17, 18}; the second spurious phase of the
sample, Ru$_2$Ge$_3$, is semiconductor and strongly diamagnetic,
exhibiting a paramagnetic contribution above 900 K due to the
structural transition~\cite{19}, from orthorhombic
(center-symmetric) to tetragonal (non-symmetrical)~\cite{20}. Thus,
in this work, we state that the low temperature magnetic,
thermodynamic and transport properties of our sample are due to the
main phase U$_4$Ru$_7$Ge$_6$ with negligible  contribution from the
secondary phases.

\begin{table}[ht]
\caption{Crystallographic parameters obtained from the Rietveld refinement for the sample.}
\begin{ruledtabular}

\begin{tabular}{lccc}

Phase & U$_4$Ru$_7$Ge$_6$ & Ru$_2$Ge$_3$ & $\gamma$-U \\ \hline \hline
Composition & 78.06\% & 20.78 \% & 1.16\% \\ \hline
Crystalline structure & cubic & orthorhombic & cubic \\   \hline
Space group & Im$\overline{3}$ & Pbcn & Im$\overline{3}$m \\ \hline
Datasheet CIF-ICSD$^*$ & 192067 & 85205 & 44392 \\ \hline
Lattice parameter &  &  &  \\
a (\AA) & 8.295813 & 11.436 & 3.504737 \\
b (\AA) & - & 9.238 & - \\
c (\AA) & - & 5.716 & - \\

\end{tabular}

\flushleft
\small{Quality of the refinement: $R_p = 19.3\%$; $R_{wp} = 16.9\%$; $R_{exp} = 8.98\%$; $\chi^2 = 3.537$; $S = 1.881$.}
\par\medskip
$^*$Crystalographic Information File - Inorganic Crystal Structure Database

\label{defaulttable}
\end{ruledtabular}
\end{table}

\section{Magnetic properties and evidence for spin-waves}

\begin{figure}[!h]
\begin{center}
\includegraphics[scale=0.32]{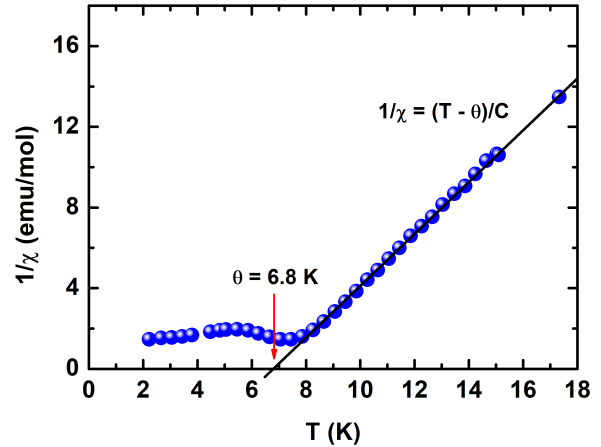}
\end{center}
\caption{(Color online) The inverse of the magnetic AC-susceptibility $\chi_{AC}(T)$ of the sample as a function of temperature for a frequency of 3 kHz, an AC-field of 1 mT and a DC field $H_{DC}=50$ mT. The peak marked with an arrow is the Curie-Weiss temperature that is very close  to the ferromagnetic transition of U$_4$Ru$_7$Ge$_6$, as indicated by the peak in the susceptibility. }
\label{Fig2}
\end{figure}

The inverse of the magnetic AC-susceptibility $\chi_{AC}(T)$ of our sample as a function of temperature is shown in Fig.~\ref{Fig2} from 2 K to 18 K. The data shows a peak at    $T_{C} = 6.8$ K that we identify as the Curie temperature, below which  U$_4$Ru$_7$Ge$_6$ becomes ferromagnetic. This coincides with the $T_{C}$ found in  Ref.~\cite{2} ($T_{C}=6.8$ K),  but is smaller than those found in Refs.~\cite{3,6} that range in the interval between $10.0-13.0$ K.
Above $T_{C}$ the susceptibility follows a Curie-Weiss law from which we extract an effective  magnetic moment of $\approx 2.14$ $\mu_B$ per Uranium atom~\cite{blundell}. This is large compared to the value obtained from the saturation magnetization  in large magnetic fields ($0.2$  $\mu_B$ per Uranium atom)~\cite{2,7}, but smaller than the value of 2.54 $\mu_B$  extracted from the Curie-Weiss behavior of  the  susceptibility at high temperatures (300 K $\lesssim$ T $\lesssim$ 500 K)~\cite{3}. It is interesting that the Curie-Weiss temperature $\theta$ indicated by an arrow  in Fig.~\ref{Fig2} is very close to the ferromagnetic critical temperature $T_{C}$ obtained from the peak in $\chi_{AC}$. This mean-field behavior is consistent with the rather large Uranium moment in a cubic structure and indicates that ferromagnetic fluctuations are important only  close to $T_{C}$.

In ferromagnetic systems, metallic or insulators, below $T_{C}$ the low temperature magnetic elementary excitations are long wave-length spin-waves with dispersion relation $\hbar \omega_k = \Delta + Dk^2$. The gap $\Delta$ may be due to the magneto-crystalline anisotropy, dipolar interactions or to the Zeeman energy if an external magnetic field $H_a$ is applied in the material. The quantity $D$ is the {\it spin-wave stiffness} of the magnetic system. For a soft ferromagnet with negligible anisotropy,  in zero field and in the temperature range of our experiments, the gap in general can be neglected and the spectrum is purely quadratic in the wave-vector $k$. Notice that in this case, the mode $\omega_{k=0}=0$ is the Goldstone mode of the rotational invariant system~\cite{02}.

\begin{figure}[!h]
\begin{center}
\includegraphics[scale=0.34]{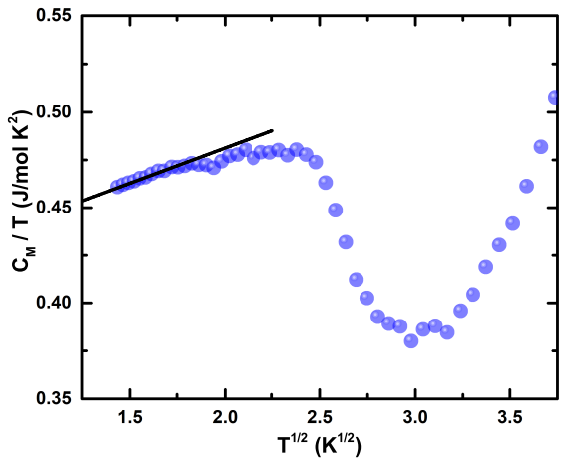}
\end{center}
\caption{(Color online) Molar specific heat  plotted as $C_M/T$ versus $T^{1/2}$ to put in evidence a $T^{3/2}$ contribution due to ferromagnetic magnons. The extrapolation of the line gives a linear temperature contribution with a coefficient $\gamma_0=0.407$ J/mol K$^2$. The coefficient of the $T^{3/2}$ term obtained from the inclination of the line turns out to be  $\delta= 0.038$ J/mol K$^{5/2}$. }
\label{Fig3}
\end{figure}

Fig.~\ref{Fig3}  shows the low temperature behavior of $C_M/T$, the molar specific heat divided by temperature,   of U$_4$Ru$_7$Ge$_6$.  We took into account that only 78\% of the total molecular weight is due to the main phase. The data is plotted in such a way to put in evidence a contribution proportional to $T^{3/2}$ associated with gapless ferromagnetic magnons with a quadratic dispersion. There is also a large linear temperature dependent term that is due to the heavy quasi-particles of the metallic U$_4$Ru$_7$Ge$_6$ compound. A phonon contribution to the specific heat  becomes apparent in a plot of $C_M/T$ versus $T^2$, where a linear behavior is observed in the temperature interval from $\approx $ 15 K  to 30 K. The Debye temperature obtained from the inclination of this line is $\Theta_D =276$ K~\cite{2}.  The data shows that this phonon contribution can be safely neglected  in the temperature region below $T_{C}$. The coefficient of the linear temperature dependent term of the specific heat obtained from the fit in Fig.~\ref{Fig3} with $C_M/T=\gamma_0 + \delta T^{3/2}$ is $\gamma_0= 102$ mJ/mol U K$^2$,  similar to that given in the literature for this material~\cite{2,6,7}. The coefficient of the spin-wave contribution obtained from this  fit  is $\delta= 0.038$ J/mol K$^{5/2}$. In  spin-wave theory,  for a cubic ferromagnetic system with gapless magnon excitations, the expression for  the contribution of these modes to the low temperature specific heat per unit volume is calculated as~\cite{7.1},
\begin{equation}
\label{eq1}
C_V=\frac{1}{V} \left( \frac{\partial E}{\partial T} \right)_V= \frac{15}{4}  \zeta(\frac{5}{2}) \left( \frac{1}{4 \pi D} \right)^{3/2} k_B^{5/2} T^{3/2}.
\end{equation}
This expression allows to obtain the spin-wave stiffness $D$ from the coefficient of the $T^{3/2}$ term of the specific heat in the experimental data as shown in Fig.~\ref{Fig3}. Notice that $C_M = C_V \mathcal{V}_M$ with $C_V$ given by Eq.~\ref{eq1} and $\mathcal{V}_M$ the molar volume.  We get, using the value of $\delta$ above, $D=32 \pm 1$ meV \AA$^{2}$. The error here is mainly due to uncertainty in the volume of the sample.
\begin{figure}[!h]
\begin{center}
\includegraphics[scale=0.33]{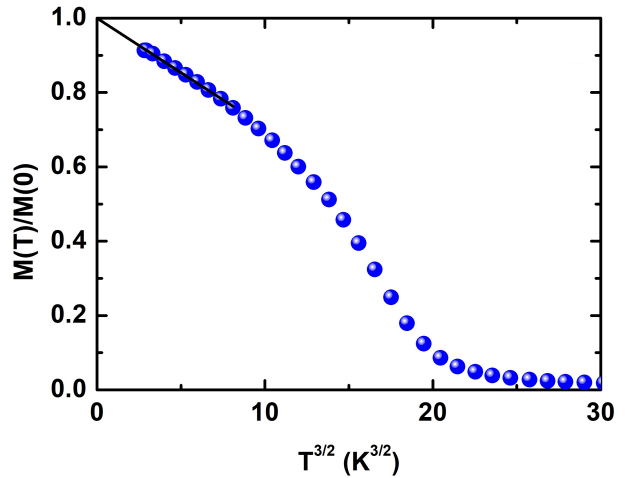}
\end{center}
\caption{(Color online) Normalized low temperature magnetization of the ferromagnet U$_4$Ru$_7$Ge$_6$ versus $T^{3/2}$ measured in a field cooled of 10 mT. At  low temperatures, the magnetization decreases with a Bloch $T^{3/2}$ law. We get for the coefficient in Eq.~\ref{blochm}, B=0.030 K$^{-3/2}$. }
\label{Fig4}
\end{figure}

Fig.~\ref{Fig4} shows the normalized low temperature magnetization of U$_4$Ru$_7$Ge$_6$ as a function of $T^{3/2}$. The linear behaviour of  the  magnetization in this plot implies that at low temperatures it decreases according to a Bloch's $T^{3/2}$ law. This is a clear signature that this decrease is due to the thermal excitation of ferromagnetic magnons~\cite{7.1}. Bloch's law yields,
\begin{equation}
\label{blochm}
M(T)/M(0)=1 - B T^{3/2}
\end{equation}
at low temperatures~\cite{7.1}. The coefficient $B$ is related to the spin-wave stiffness by~\cite{7.1},
\begin{equation}
B= \frac{\zeta(3/2)g \mu_B}{M(0)}\left(\frac{k_B}{4 \pi D}\right)^{3/2}.
\label{bloch}
\end{equation}
Using the experimental value of $B$ obtained from Fig.~\ref{Fig4} in Eq.~\ref{bloch}, we find $D=27 \pm 1$ meV \AA$^2$, where the error comes from the uncertainty on the volume of the sample. In Eq.~\ref{blochm}, M(0)= 6.1 x 10$^6$ emu/m$^3$

The values for the spin-wave stiffness obtained above from the low temperature specific heat and magnetization  measurements, $D=32 \pm 1.0$ meV \AA$^{2}$ and $D=27 \pm 1.0$ meV \AA$^{2}$ respectively, are consistent and of the expected order of magnitude for a soft metallic ferromagnet with a Curie temperature of $T_{C} \approx 10$ K. For example, in ferromagnetic Ni with $T_{C} \approx 631$ K~\cite{7.1} the experimentally obtained spin-wave stiffness ranges from $D \approx 422$ meV \AA$^2$ to $D \approx 555$ meV \AA$^2$~\cite{7.2}, if extracted from Bloch's law or measured directly by neutron scattering, respectively.
These results  strongly support the idea that ferromagnetic magnons play an important role in the thermodynamic properties of the cubic ferromagnetic  U$_4$Ru$_7$Ge$_6$ below its Curie temperature.

\section{Transport properties}

In ferromagnetic metals, the scattering of conduction electrons by ferromagnetic spin-waves gives rise to a $T^2$ temperature dependent contribution to the electrical resistivity of these materials~\cite{7.3}. This $T^2$ contribution has indeed been found in a previous study of U$_4$Ru$_7$Ge$_6$~\cite{6}, but it has been attributed to electron-electron scattering as in strongly correlated paramagnetic metals. However, in a long-range ordered ferromagnetic metal, with polarized bands, and gapless spin-wave modes, the main scattering is due to these elementary excitations. In fact, this is the main form electron-electron scattering assumes in an itinerant ferromagnetic metal. Here, we  give evidence that in ferromagnetic U$_4$Ru$_7$Ge$_6$ substantial part of the $T^2$ term in its resistivity is due to electron-magnon scattering. This contribution to the resistivity is given by~\cite{7.3},
\begin{equation}
\rho=A T^2=\frac{32}{3} \zeta(2) \pi^2 \rho_0 \left(\frac{\Delta E}{E_F}\right) \left( \frac{k_B T}{Dk_F^2}\right)^2
\label{rsw}
\end{equation}
where $\rho_0$ is a constant with units of resistivity, $k_F$ and $E_F$, the Fermi wave-vector and Fermi energy of the $f$-electrons, respectively. The quantity  $(\Delta E/E_F)=2q_{max}/k_F$ where $q_{max}$ is the maximum wave-vector for which the quadratic spin-wave dispersion relation,  $\hbar \omega=Dk^2$, holds. The spin-wave stiffness appears in the denominator of this equation, such that, the softer the magnon modes the larger is this contribution to the resistivity. At a FQCP, where the spin-wave stiffness vanishes, the resistivity of the metal is given by\cite{7.3},
\begin{equation}
\rho=64 \pi \rho_0 \Gamma(\frac{8}{3}) \zeta(\frac{5}{3})  \left( \frac{3 \pi k_B T}{E_F}\right)^{5/3}.
\label{rfqcp}
\end{equation}
In the next section, we will present results for the electric resistivity of our sample as a function of temperature for different applied pressures and magnetic fields.
As pressure increases, the ferromagnetic Curie temperature vanishes smoothly at a FQCP at a critical pressure, $P_c=1.7 \pm 0.02$ GPa. The resistivity curves we obtain present no hysteresis for any pressure. We find no evidence of a behavior  that could indicate a first order transition as the Curie temperature of the sample is reduced and made to vanish.
As we accompany the variation of the coefficient of $T^2$ term with increasing pressure, we observe a smooth increase of this coefficient that we attribute entirely, according to Eq.~\ref{rsw},  to a decrease of the spin-wave stiffness as the FQCP is approached. At the critical pressure the resistivity follows a $T^{5/3}$ behavior in agreement with Eq.~\ref{rfqcp}.

\subsection{Pressure experiments}

\begin{figure}[!h]
\begin{center}
\includegraphics[scale=0.35]{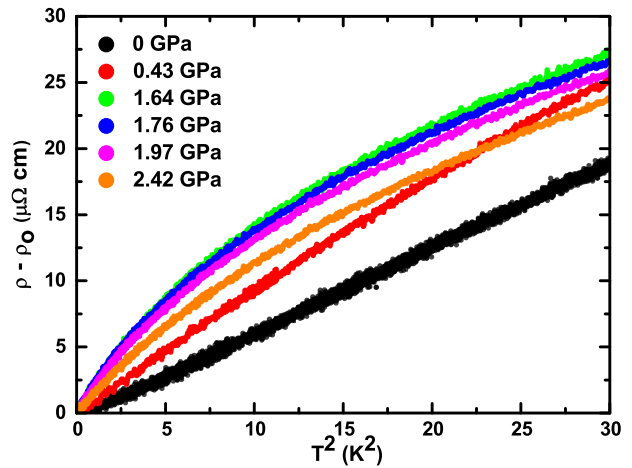}
\end{center}
\caption{(Color online) Resistivity versus temperature of the ferromagnet U$_4$Ru$_7$Ge$_6$ for different applied pressures above and below the critical pressure. The residual resistivity and the coefficients of the low temperature $T^2$ terms for different pressures are shown in Fig.~\ref{Fig7}.  }
\label{Fig5}
\end{figure}

Fig.~\ref{Fig5} shows the low temperature behavior of the electrical resistivity of U$_4$Ru$_7$Ge$_6$ as a function of pressure. All pressure experiments were carried out in zero external magnetic field.
\begin{figure}[!ht]
\begin{center}
\includegraphics[scale=0.36]{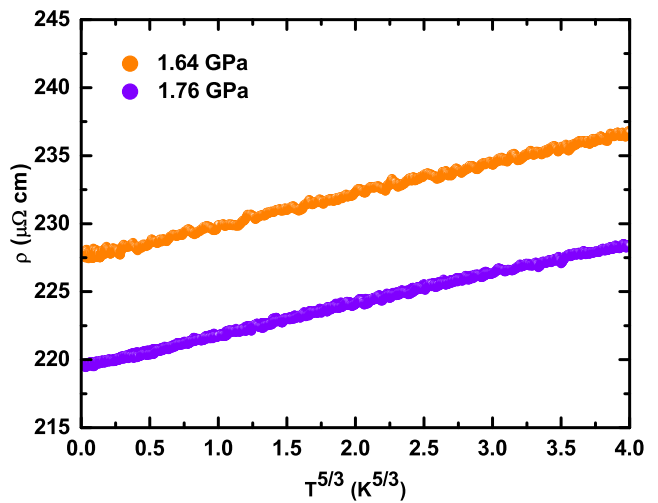}
\end{center}
\caption{(Color online) Low temperature resistivity  of the ferromagnet U$_4$Ru$_7$Ge$_6$ for  pressures very close to the pressure where ferromagnetism is suppressed. The plot puts in evidence a $T^{5/3}$ power law behavior expected for an itinerant $3d$ ferromagnet   at a FQCP (see Eq.~\ref{rfqcp}).  }
\label{Fig6}
\end{figure}
\begin{figure}[!ht]
\begin{center}
\includegraphics[scale=0.33]{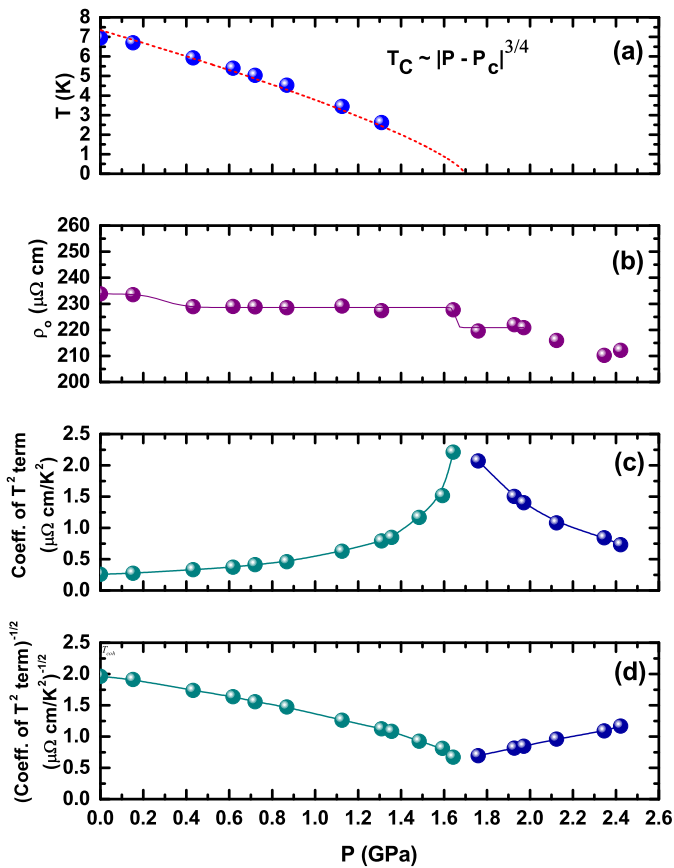}
\end{center}
\caption{(Color online) Parameters extracted from the temperature dependent resistivity curves for different pressures. Upper panel, the Curie temperature obtained from the second derivative of the resistivity (see text).  The dashed line  corresponds to a fitting with the expression $T_C \propto |P_c - P|^{\psi}$ with $\psi=3/4=0.75$, the expected shift exponent for a 3d itinerant ferromagnet (see text), and $P_c=1.7$ GPa.   The panel below  shows the residual resistivity as a function of pressure. The lower two panels refer to the coefficients of the $T^2$ term in the resistivity, above and below $P_c$.  }
\label{Fig7}
\end{figure}
\begin{figure}[!ht]
\begin{center}
\includegraphics[scale=0.36]{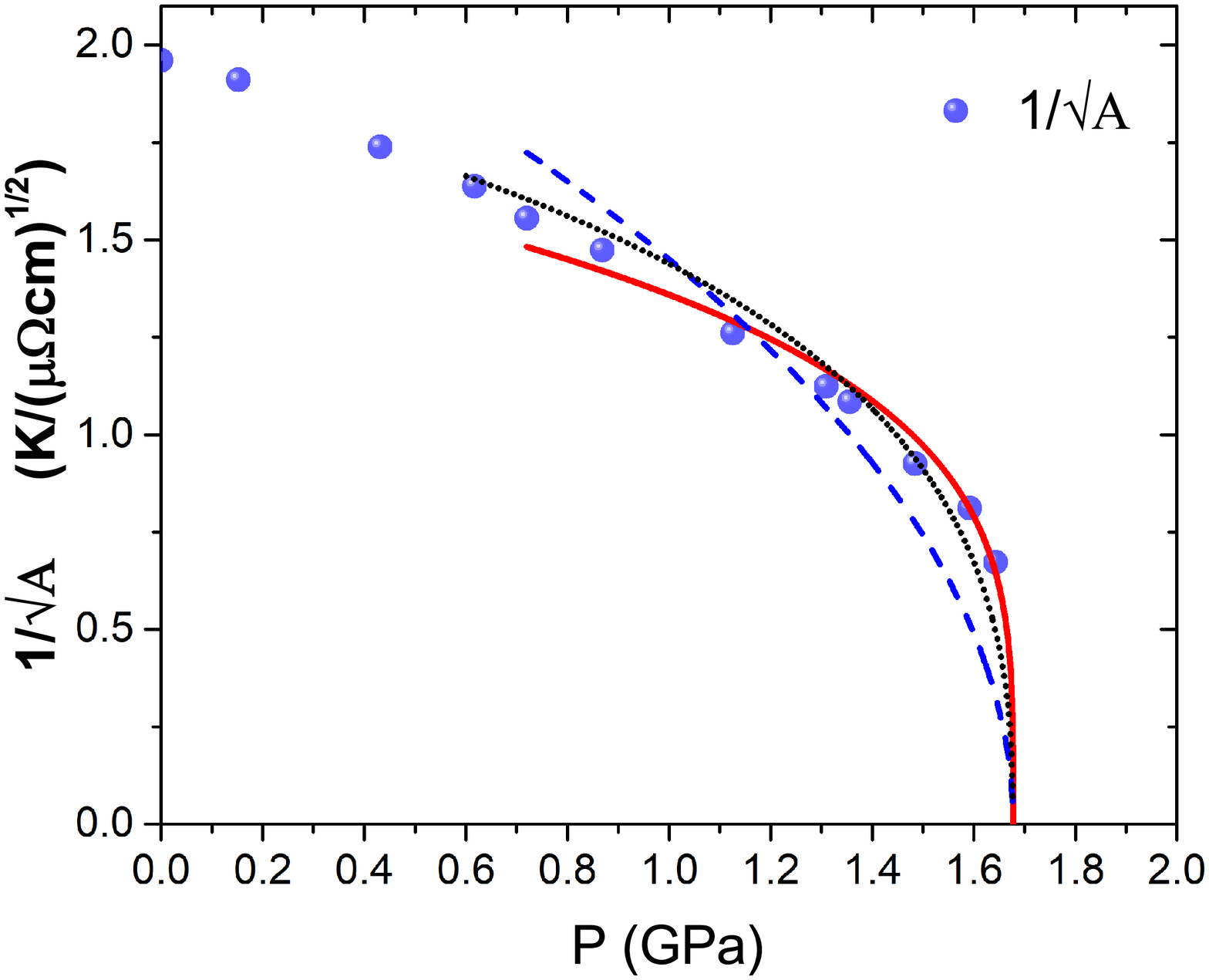}
\end{center}
\caption{(Color online) Pressure dependence of the quantity $1/\sqrt{A}$, that mimics that of the spin-wave stiffness $D$. Dashed line is the mean-field prediction $D \propto m$, while the full and dotted  lines are best fittings using Eqs.~\ref{dsw} and~\ref{csw}, respectively. In every case $P_c=1.64$ GPa.  }
\label{Fig8}
\end{figure}
The electrical resistivity behaves smoothly, with no detectable hysteresis for all pressures of the experiments. At very low temperatures, it presents a $T^2$ dependence, both above and below the critical pressure $P_c \approx 1.7$ GPa and is well described by, $\rho=\rho_0 + A(P) T^2$. The  exception is for pressures very close to $P_c$ where $\rho(T) \propto T^{5/3}$, as shown in Fig.~\ref{Fig6}. This is the expected power law behavior for a ferromagnetic metallic system close to a FQCP (see Eq.~\ref{rfqcp}).

In Fig.~\ref{Fig7}, we show physical parameters obtained from the resistivity data as a function of pressure.
\begin{itemize}
\item
In principle, the observation of a clear cut-bend (or kink) in the $T$-dependence of electrical resistivity is an indication of an onset of magnetic ordering. The precise determination of $T_{C}$  is obtained from the second derivative of the smoothed electrical resistivity data~\cite{julio}.
In the upper panel of Fig.~\ref{Fig7}, we have plotted the $T_{C}$  obtained in this way  and  draw through these points a curve from a fit using the expected power law behavior for an itinerant FQCP, $T_{C} \propto |P_c-P|^{\psi}$, where the {\it shift} exponent~\cite{02} $\psi = z/(d+z-2)= 3/4$, since the dynamic exponent is $z=3$ in this case~\cite{02}. The curve gives a reasonable description of the pressure dependent Curie temperatures.

\item The next panel shows the pressure dependence of the residual resistivity. This is nearly constant in the ferromagnetic phase, with a small drop close to the critical pressure.

\item The last two panels refer to the pressure dependence of the coefficient of the $T^2$ term of the resistivity. They rise on both sides of the phase diagram as the critical pressure is approached from below and above in a non-symmetric fashion. In the paramagnetic phase, above $P_c$, this term is due to scattering by paramagnons and its  coefficient is proportional to the square of the inverse of the coherence temperature~\cite{021}, $T_{coh} \propto |P-P_c|^{\nu z}$, with $\nu z=3/2$ for a three dimensional  itinerant ferromagnetic system~\cite{021}. As can be seen in the last two panels, we do not have enough data for $T_{coh}$ close to the critical pressure to be able to determine its power law dependence with the distance from criticality. Sufficiently far from $P_c$, $T_{coh}$ depends linearly on this distance, which suggests local quantum critical behavior~\cite{02}.

\item For pressures below $P_c$,  in the ferromagnetic phase, according to Eq.~\ref{rsw} the coefficient of the $T^2$ term in the resistivity is related to spin-wave stiffness $D$, $A(P) \propto 1/D^2$.
In itinerant 3d ferromagnets, the coupling of the order parameter to particle-hole excitations give rise to a non-analytic behavior of the spin-wave stiffness  as a function of the  magnetization $m$~\cite{belitz}. For a disordered quantum itinerant 3d ferromagnet,
\begin{equation}
D(m \rightarrow 0)=c_3 m [m^{-1/2} + O(1)]
\label{dsw}
\end{equation}
while for the clean system,
\begin{equation}
D(m \rightarrow 0)=\tilde{c}_3 m [\ln(1/m) + O(1)]
\label{csw}
\end{equation}
where $c_3$ and $\tilde{c}_3$ are positive constants~\cite{belitz}.
In a 3d quantum metallic ferromagnet the magnetization vanishes close to the FQCP as $m \propto |P_c-P|^{\beta}$ with a mean-field exponent $\beta=1/2$.
In Fig.~\ref{Fig8} we plot the pressure dependence of the quantity $1/\sqrt{A}$ that mimics that of the spin-wave stiffness for pressures approaching the FQCP. We compare the simple mean-field result $D \propto m \propto \sqrt{(P_c-P)}$ with the results of Eqs.~\ref{dsw} and ~\ref{csw}. The results for the clean and disordered ferromagnet are rather similar, but clearly they give a better description of the data than the simple mean-field.

\end{itemize}

\section{Effect of an external magnetic field}

In this section, we study the effect of an applied magnetic field on the thermodynamic and transport properties of our sample at ambient pressure.
In ferromagnets, a magnetic field is the conjugate of the order parameter and destroys the thermodynamic phase transition. This is different from the antiferromagnet where a uniform magnetic field just shifts the transition. The low temperature magnetic excitations of the ferromagnet in an external magnetic field are still magnons, but they become partially quenched by a {\it Zeeman gap} due to the coupling of the magnetic moments to the field.
\begin{figure}[!ht]
\begin{center}
\includegraphics[scale=0.36]{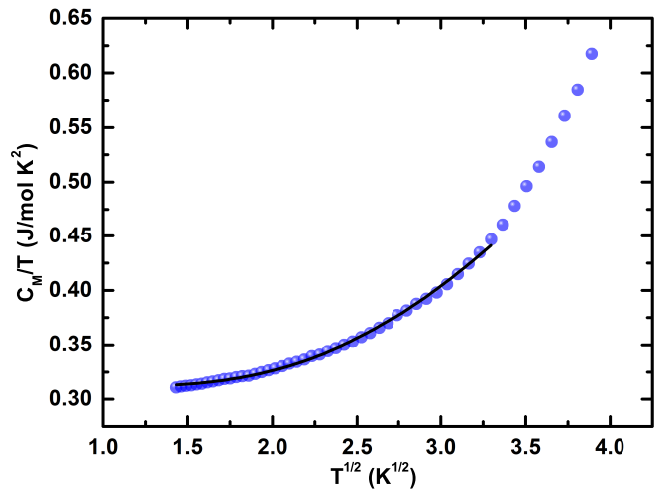}
\end{center}
\caption{(Color online) Specific heat of U$_4$Ru$_7$Ge$_6$  as a function of temperature in an external magnetic field $H_a=7$ T. The fitting curve $C_M/T =\gamma + b \sqrt{T} e^{-\Delta/T}$ includes an exponential term that accounts for the {\it freezing} of the magnons by the external field. The parameters $\gamma= 310$ mJ/mol K$^2$ and $\Delta=9.1$ K. }
\label{Fig9}
\end{figure}
\begin{figure}[!ht]
\begin{center}
\includegraphics[scale=0.33]{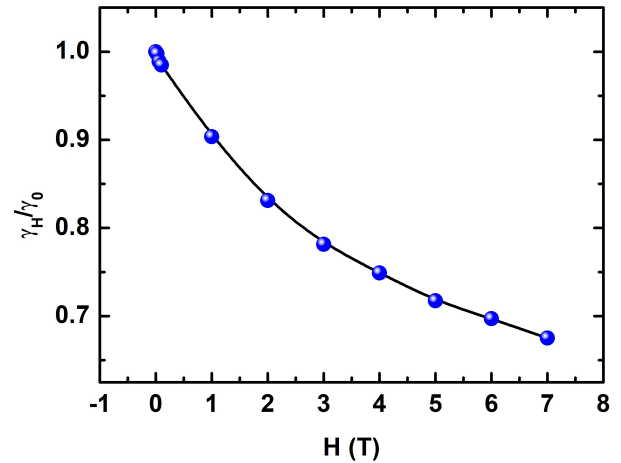}
\end{center}
\caption{(Color online) The coefficient of the linear term in the specific heat as a function of the external magnetic field.  The line is a guide to the eyes. }
\label{Fig10}
\end{figure}
\begin{figure}[!ht]
\begin{center}
\includegraphics[scale=0.32]{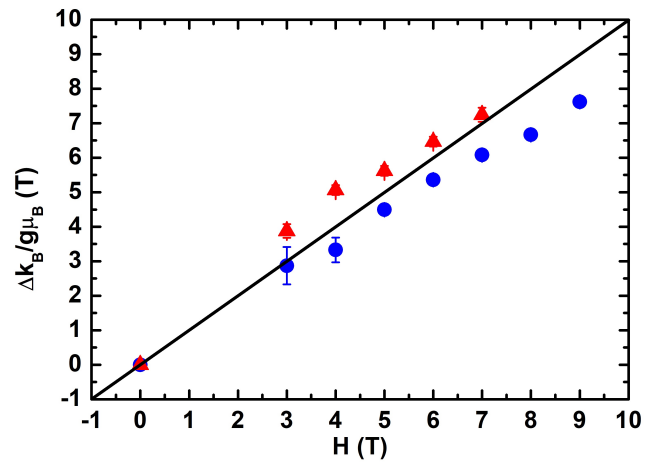}
\end{center}
\caption{(Color online) The spin-wave gaps extracted from the specific heat data (triangles) and from the resistivity (circles) using the expressions in the text. The straight line is the result expected from the simplest spin-wave theory (see text).}
\label{Fig11}
\end{figure}
\begin{figure}[!ht]
\begin{center}
\includegraphics[scale=0.39]{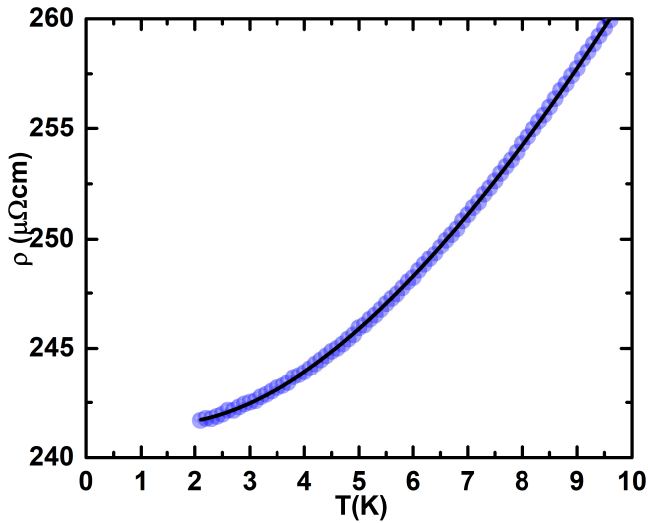}
\end{center}
\caption{(Color online) Resistivity of our sample as a function of temperature in an external magnetic field $H_a=7$ T. The fitting  curve $\rho=\rho_0 + b_r T^2 e^{-\Delta/T} + c_r T e^{-\Delta/T}$ (black) takes into account  the quenching of the magnons by the magnetic field that suppresses the electron-magnon scattering at very low temperatures (see text). The gap for this field is $\Delta=6.5$ K.   }
\label{Fig12}
\end{figure}
This reduces the influence and contribution of spin-waves to the low temperature properties, i.e., for $k_B T < \Delta$, where $\Delta$ is the Zeeman gap.

In Fig.~\ref{Fig9} we show the specific heat as a function of temperature in an applied field of 7 T. The  low temperature specific heat is well fitted by the expression $C_M/T = \gamma +b \sqrt{T} \exp(-\Delta/T)$. The exponential term takes into account the quenching of the magnons by the Zeeman gap $\Delta = (g \mu_B S H_a)/k_B$, expressed here in temperature scale. From this fit we can determine the coefficient of the linear term $\gamma(H_a)$ and the Zeeman gap $\Delta(H_a)$ for several values of the external magnetic field, as shown in Figs.~\ref{Fig10} and~\ref{Fig11}, respectively.
Fig.~\ref{Fig10} shows that the coefficient of the linear term  in the specific heat is reduced as the external field is applied.  The simplest interpretation for this effect is that the Zeeman splitting of the polarized bands causes a decrease in the density of states at the Fermi level~\cite{stewart}. This behavior of $\gamma(H_a)$ is quite distinct from that in antiferromagnet heavy fermions~\cite{sheinkin},  as expected from the different roles of $H_a$ in these systems.

The suppression of the magnons by the magnetic field also decreases the low temperature electrical resistivity due to a partial freezing of the  electron-magnon scattering. The low temperature electrical resistivity shown in Fig.~\ref{Fig12} for $H_a=7$ T is well described by the expression~\cite{rhdelta}
\begin{equation}
\rho=\rho_0 + a \Delta T e^{-\Delta/ T}\left(1 + 2 \frac{ T}{\Delta}\right).
\end{equation}
The gap $\Delta$ (units of temperature) extracted from the resistivity data for several values of the applied magnetic field is shown in Fig.~\ref{Fig11}.
The gaps obtained from the transport and thermodynamic data are in satisfactory agreement. The straight line in this plot show the expected value for the Zeeman gap in the simplest (non self-consistent) spin-wave theory~\cite{21}.

\begin{figure}[!ht]
\begin{center}
\includegraphics[scale=0.37]{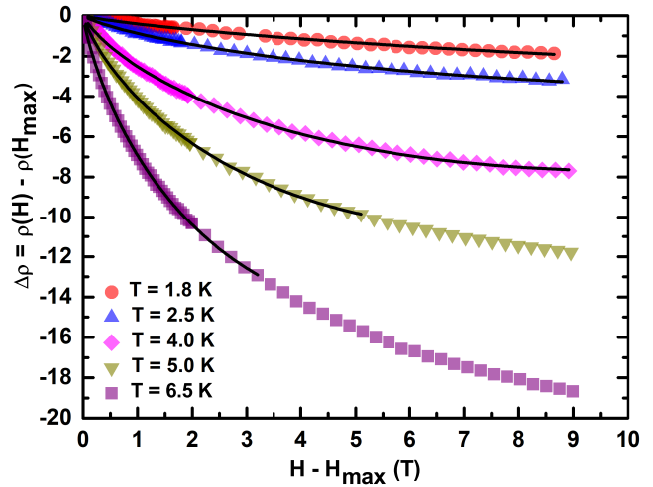}
\end{center}
\caption{(Color online) Magneto-resistance of our sample as a function of of magnetic field for several fixed temperatures. The lines are two-parameters fittings ($a$ and $b$) using the expression $\Delta \rho= a(H-H_{max}) \ln(H-H_{max})/b$ as given in the text~\cite{mamy}.   }
\label{Fig13}
\end{figure}

For completeness, we   now present magneto-resistance results at ambient pressure for our sample. Resistivity is measured as a function of magnetic field for fixed temperatures below the Curie temperature. For small fields, $H_a \ll 1$ T and very low temperatures, the magnetoresistance is positive reaches a maximum at $H_{max}$ and then decreases almost linearly with field for $H_a > H_{max}$. Magneto-resistance of  multi-band ferromagnetic metals, like transition metals, has been intensively studied both experimentally and theoretically~\cite{mamy}. Raquet {\it et al}.~\cite{mamy} have shown that for systems with a light $c$-{\it band} of conduction electrons and a heavy $f$-{\it band} of quasi-particles,  intra-band scattering in the conduction c-band can be neglected. Notice that due to the strong $c-f$ hybridization in U$_4$Ru$_7$Ge$_6$~\cite{7} the bands have a hybrid character and those referred above are in fact are {\it mostly c-band} and {\it mostly f-band}.  Considering electron-magnon scattering, which  involves  intra-band $f-f$ and inter-band $c-f$ spin flip process,  Raquet {\it et al}.~\cite{mamy} have shown that in the presence of a magnetic field, the magneto-resistance  roughly follows a $H_a \ln H_a$ dependence for temperatures above approximately $T_{C}/5$.
In Fig.~\ref{Fig13} we plot the magneto-resistance defined as $\Delta \rho = \rho(H_a,T)-\rho(H_{max},T)$ as a function of $H_a - H_{max}$ for different fixed temperatures. $H_{max}$ is the value of the magnetic field for which the magnetoresistance reaches a maximum before starts to decrease. We attribute the positive magneto-resistance at low fields and temperatures to the existence of domain walls  that are eventually removed at $H_{max}$.
Fig.~\ref{Fig13} also shows the fittings using the simple logarithmic law obtained in Ref.~\cite{mamy}, $\Delta \rho= a \delta H \ln (\delta H/b)$, with $\delta H=H-H_{max}$. It is clear it gives a good description of our data.

\section{Conclusions}

The study of systems close to quantum criticality is an exciting area of research. In the case of itinerant ferromagnets driven to a magnetic instability,  there are theoretical and experimental evidences that quantum critical behaviour is avoided and a first order transition occurs before the FQCP is reached~\cite{fqcp,45}. 
In this work we present a thorough investigation of the ferromagnetic heavy fermion system U$_4$Ru$_7$Ge$_6$ as it is driven to the paramagnetic state under applied pressure. The results of the transport properties under pressure  show no sign of a discontinuous behavior as $T_{C}$ is reduced. The resistivity curves present no hysteresis   effects for any pressure, before and after ferromagnetism is suppressed. Disorder is certainly present in our system, as evidenced by its high residual resistivity. It is possible that the continuous behavior is due to its influence since it substantially modifies the properties  of our sample when compared to single crystals~\cite{7}. However, disorder is not sufficiently strong to give rise to localization effects or Griffith's singularities. On the contrary, our sample presents many of the properties expected for an itinerant clean system,  as the $T^{5/3}$ temperature dependence of the resistivity at its FQCP.

U$_4$Ru$_7$Ge$_6$ is a unique Uranium compound with negligible anisotropy. 
This implies that  spin-waves, the elementary excitations of a ferromagnetic metal can be easily excited and play a fundamental role in the thermodynamic and transport properties of this system at low temperatures. We have shown that as the FQCP is approached with increasing pressure the spin-wave stiffness softens and we obtain its quantum critical behavior.
It would be very interesting to measure directly the spin-wave stiffness of U$_4$Ru$_7$Ge$_6$ by inelastic neutron scattering and compare it to the results we obtained. Specially interesting would be to observe with neutrons the softening of the magnons with increasing pressure.

Although our results are in close  agreement with those of Refs.~\cite{2} and ~\cite{6} on the same system, they differ from those obtained in single crystals~\cite{7} where a reorientation of the magnetization  inside the ferromagnetic phase is observed.

In summary our results on the ferromagnetic U$_4$Ru$_7$Ge$_6$ provide strong evidence for the existence of a pressure induced ferro-para quantum phase transition in this system that is accompanied by a softening of the elementary excitations of the ordered phase.  Its quantum critical behavior  shares many features with that expected for a clean, itinerant FQCP.

\acknowledgments The authors M.P.N., M.A.C., A.L., A.L., D.C.F.,
J.L.J., C.E., J.F.O., E.B.S. and M.B.F. would like to thank the
Brazilian Agencies CAPES, CNPq and FAPERJ for partial financial
support. We thank  Prof. Dr. Renato Bastos Guimar\~aes for using the
X-Ray Diffraction Laboratory (LDRX) at the Instituto de F\'{\i}sica,
Universidade Federal Fluminense (IF-UFF).

\end{document}